\documentclass[aps,preprint,nofootinbib,preprintnumbers,eqsecnum]{revtex4-1}

\usepackage{color}

\newcommand{\nb}[1]{\color{blue}}

\newcommand{\hl}[1]{\color{magenta}}

\usepackage[
	  pagebackref=false,
	  colorlinks=true,
      linkcolor=blue,
      urlcolor=blue,
      filecolor=black,
      citecolor=red,
      pdfstartview=FitV,
      pdftitle={},
        pdfauthor={Paolo Glorioso, Michael Crossley, Hong Liu},
        pdfsubject={},
        pdfkeywords={},
        pdfpagemode=None,
        bookmarksopen=true
      ]{hyperref}

\usepackage[normalem]{ulem}
\usepackage{amsmath}
\usepackage{enumerate}
\usepackage{amsfonts}
\usepackage{epsfig}
\usepackage{mathbbol}

\def\Tr{\mathop{\rm Tr}}

\newcommand\half{{\ensuremath{\frac{1}{2}}}}
\newcommand\p{\ensuremath{\partial}}

\newcommand{\be}{\begin{equation}}
\newcommand{\ee}{\end{equation}}
\newcommand{\bea}{\begin{eqnarray}}
\newcommand{\eea}{\end{eqnarray}}
\newcommand{\bega}{\begin{gather}}
\newcommand{\ega}{\end{gather}}

\newcommand{\bi}{\begin{itemize}}
\newcommand{\ei}{\end{itemize}}
\newcommand{\ben}{\begin{enumerate}}
\newcommand{\een}{\end{enumerate}}
\newcommand{\bca}{\begin{cases}}
\newcommand{\eca}{\end{cases}}
\newcommand{\bln}{\begin{align}}
\newcommand{\eln}{\end{align}}
\newcommand{\bst}{\begin{split}}
\newcommand{\est}{\end{split}}
\def\ie{\begin{equation}\begin{aligned}}
\def\fe{\end{aligned}\end{equation}}
\newcommand{\bma}{\le(\begin{matrix}}
\newcommand{\ema}{\end{matrix}\ri)}

\newcommand\al{{\alpha}}
\newcommand\ep{\epsilon}
\newcommand\sig{\sigma}

\newcommand\lam{\lambda}

\newcommand\om{\omega}

\newcommand\de{{\ensuremath{{\delta}}}}

\newcommand\vp{\varphi}

\newcommand\ov{\over}
\newcommand\ha{{\half}}

\def\le{\left}
\def\ri{\right}

\newcommand\sL{{\ensuremath{{\mathcal L}}}}

\newcommand\sO{{\ensuremath{{\mathcal O}}}}
\newcommand\sP{{\ensuremath{{\mathcal P}}}}

\newcommand\vx{{\vec x}}
\newcommand\vk{{\vec k}}

\begin{document}

\title{A prescription for holographic Schwinger-Keldysh contour in non-equilibrium systems}

\preprint{MIT-CTP/5095}
\preprint{EFI-18-21}

\author{Paolo Glorioso}
\affiliation{Kadanoff Center for Theoretical Physics and Enrico Fermi Institute\\
University of Chicago, Chicago, IL 60637, USA}

\author{Michael Crossley and Hong Liu}
\affiliation{Center for Theoretical Physics, \\
Massachusetts
Institute of Technology,
Cambridge, MA 02139 }

\begin{abstract}

\noindent
We develop a prescription for computing real-time correlation functions defined on a Schwinger-Keldysh contour for non-equilibrium systems using gravity. The prescription involves a new analytic continuation procedure in a black hole geometry which can be dynamical. For a system with a slowly varying horizon,  the continuation enables computation of the Schwinger-Keldysh generating functional using derivative expansion, drastically simplifying calculations. We illustrate the prescription with two-point functions for a scalar operator in  both an equilibrium state and a slowly varying non-equilibrium state. In particular, in the non-equilibrium case, we derive a spacetime-dependent local temperature from the KMS condition satisfied by the two-point functions. We then use it to derive from gravity the recently proposed non-equilibrium effective action for diffusion.


\end{abstract}

\today

\maketitle

\tableofcontents

\section{Introduction}

\begin{figure}[!h]
\begin{center}
\includegraphics[width=13cm]{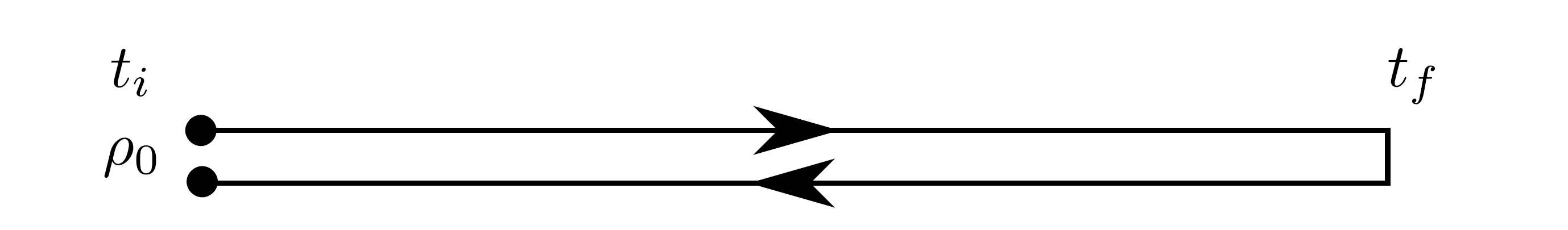}
\end{center}
\caption{A Schwinger-Keldysh contour, or often referred to as a closed time path. Operators inserted on the upper and lower segments are labeled respectively by indices $1$ and $2$.}
 \label{fig:sk}
\end{figure}

For a quantum many-body system in a state given by a density matrix $\rho_0$, various real-time correlation functions can be obtained from path integrals on a Schwinger-Keldysh contour, see Fig.~\ref{fig:sk}. More explicitly, it is often convenient to consider the generating functional
\be\label{gen0}\begin{gathered}
e^{W [\phi_{1i}, \phi_{2i}]}
= \Tr \left[\rho_0 \sP e^{ i \int d t \, (\sO_{1i} (t) \phi_{1i} (t) - \sO_{2i} (t) \phi_{2i} (t)) }\ri]\,,
\end{gathered}
\ee
where $\sO_i$ denote generic operators and $\phi_i$ their corresponding sources. $\sP$ indicates that the operators are path ordered, and subscripts $1,2$ in $\sO_{1i}$ and $\sO_{2i}$ denote the segments of the contour an operator $\sO_i$ is inserted.
The minus sign in the second term comes from the reversed time integration for the second (lower) segment.
In~\eqref{gen0} we have suppressed the spatial dependence of an operator. Correlation functions obtained from~\eqref{gen0} correspond to the full set of nonlinear response and fluctuating functions~\cite{Chou:1984es,Wang:1998wg,bernard,peterson,Lehmann:1957zz}, and thus play key roles in studies of
non-equilibrium systems.

For systems with a gravity dual, a prescription for computing~\eqref{gen0} when $\rho_0$ is a thermal equilibrium
state has been developed in~\cite{Herzog:2002pc,Son:2009vu} (see~\cite{Liu:2018crr} for a recent review). More generally, when state $\rho_0$ can be prepared by a Euclidean path integral (which includes thermal equilibrium states), one can write~\eqref{gen0} as a path
integral involving some Euclidean and some Lorentzian segments, and can obtain a corresponding gravity spacetime
by patching together different pieces. With the full path integration contour represented on the gravity side, the generating
functional~\eqref{gen0} can then be obtained using the standard procedure of integrating over the bulk fields with sources as boundary conditions~\cite{Skenderis:2008dh,Skenderis:2008dg}. This approach is conceptually straightforward, but in practice tedious to carry out even for a thermal equilibrium computation. Also general non-equilibrium state cannot be prepared by a Euclidean path integral. See also~\cite{CaronHuot:2011dr,Chesler:2011ds,Botta-Cantcheff:2018brv} for other discussions of real-time correlation functions in non-equilibrium contexts.

In this paper we propose a new prescription for computing~\eqref{gen0} for a non-equilibrium state whose
dual gravity solution involves a dynamical horizon which is analytic.\footnote{This prescription was already briefly advertised in~\cite{Liu:2018crr}.}  The prescription involves a simple analytic continuation of the black hole geometry.  Even for a thermal equilibrium state, the prescription has a number of advantages over previous approaches. For example,
for external sources which are slowly varying, it allows one to perform derivative expansion straightforwardly, which is not possible using the analytic continuation approach of~\cite{Herzog:2002pc,Son:2009vu}. Operationally it is much simpler than the approach of~\cite{Skenderis:2008dh,Skenderis:2008dg}. We illustrate the approach using the example of scalar two-point functions in  both an equilibrium state and a slowly varying non-equilibrium state. In particular, in the non-equilibrium case, we derive a spacetime-dependent local temperature from the KMS condition satisfied by the two-point functions.
The simplicity of the approach also enables us to derive from gravity
the effective action for diffusion formulated in~\cite{CGL}, which requires isolating the gapless diffusion mode from~\eqref{gen0} when $\sO$ is given by a conserved current.\footnote{See~\cite{Glorioso:2018wxw} for a review. See also~\cite{Dubovsky:2011sj} for a non-dissipative formulation, \cite{Harder:2015nxa} for a preliminary dissipative formulation, and~\cite{yarom,GaoL} for superspace formulations. See also \cite{Chen-Lin:2018kfl} for a recent interesting application.}

The plan of the paper is as follows. In Sec.~\ref{sec:2} we introduce the prescription. In Sec.~\ref{sec:3} we apply the prescription to obtain the large-distance and long-time behavior of two-point functions of a scalar operator using derivative expansion, both in an equilibrium state and in a non-equilibrium state with a slowly varying horizon. In Sec.~\ref{sec:vec} we apply the prescription to derive an effective action for diffusion. We conclude in Sec.~\ref{sec:5} with a discussion of future directions.

\emph{Note:} As this work was nearing completion we became aware of \cite{deBoer:2018qqm} which also derived the effective action for diffusion from holography.

\section{A new analytic continuation procedure} \label{sec:2}

We will now discuss a proposal for the gravity description of the Schwinger-Keldysh contour of Fig.~\ref{fig:sk}
for a general time-dependent gravity geometry with an analytic horizon.

Consider a time-dependent gravity solution
in Eddington-Finkelstein coordinates, 
\be\label{metric}
ds^2=-f(r, x^\mu) dv^2 +2dr dv+\lambda_{ij}(r, x^\mu) dx^i dx^j\ ,
\ee
with $x^\mu = (v, x^i)$. Throughout the paper we will use $0$ to denote $v$-indices. We assume the metric to be asymptotically AdS, with boundary lying at $r=\infty$,
\be
\label{metric1} f(r\to\infty)\to r^2,\qquad \lambda_{ij}(r\to\infty)\to\delta_{ij}r^2\ .
\ee
We also assume $f$ and $\lam_{ij}$ are smooth functions of $r$ and $x^\mu$, with $f(r, x^\mu)$ having a simple zero at some $r_h (x^\mu)$. In this section we shall not need to specify  the form of $f$ and $\lambda_{ij}$.

\begin{figure}[!h]
\begin{center}
\includegraphics[width=12.5cm]{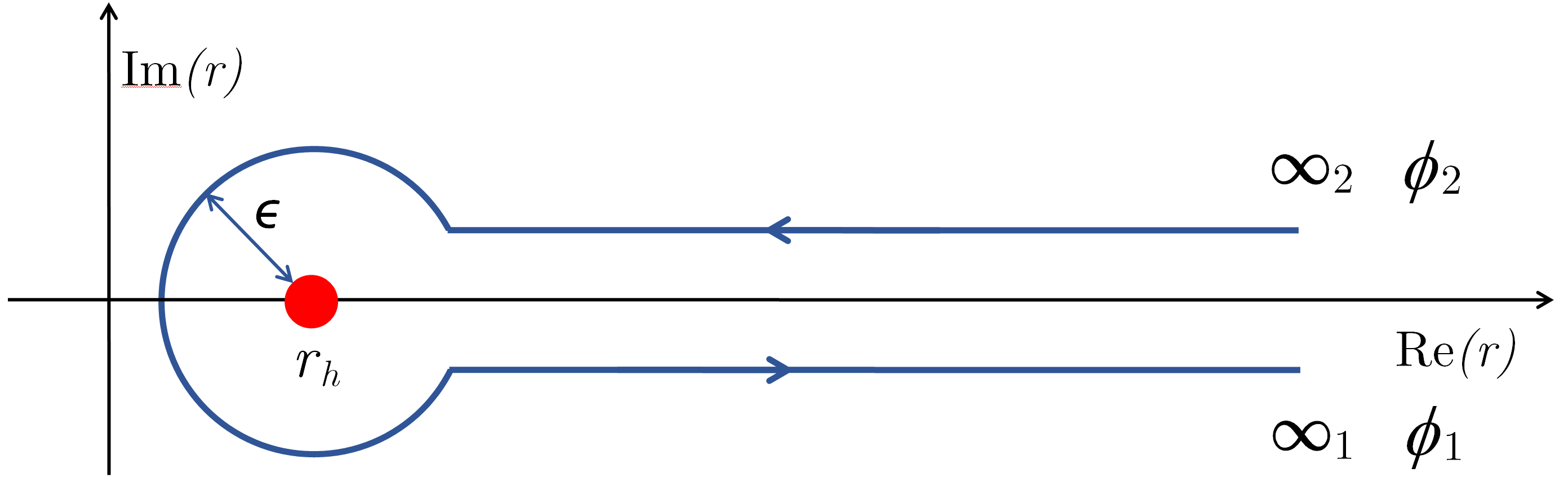}
 \end{center}
\caption{The complexified gravity spacetime corresponding to the Schwinger-Keldysh contour in Fig.~\ref{fig:sk}. The two segments of gravity spacetime are identified with the two segments of the Schwinger-Keldysh contour.
It should be understood that in the above plot the circle around $r_h$ is infinitesimal and
the two segments along the real $r$-axis have infinitesimal imaginary parts. The two boundaries are denoted respectively as $\infty_{1,2}$ with possible sources $\phi_{1,2}$. }
 \label{fig:contour}
\end{figure}

Now to describe the Schwinger-Keldysh contour in Fig.~\ref{fig:sk} on the gravity side, we
treat the radial coordinate $r$ as a complex variable, and analytically continue $r$
around $r_h$ as indicated in Fig.~\ref{fig:contour}.  The part of the contour below the real $r$-axis is identified with the first (upper) segment of the Schwinger-Keldysh contour of Fig.~\ref{fig:sk}, while the part above the real $r$-axis is identified with the second (lower) Schwinger-Keldysh segment. The two slices are connected by a circle going around $r_h$ in a counterclockwise manner. The radius $\ep$ of the circle is taken to be infinitesimal.
The arrows in Fig.~\ref{fig:contour} denote the orientations of the $r$-direction, where the lower segment has the standard orientation while the upper segment has the opposite orientation. The metric~\eqref{metric} should now be understood as that on the full $r$-contour. For the full complexified spacetime to have a single orientation, the reversal of $r$-orientation effectively reverses the orientation of the $v$ direction in the second copy. Note that one does no analytic continuation in Eddington-Finkelstein time $v$, which is already regular.

For an external black hole which describes a thermal equilibrium state, the above prescription is equivalent to
that of~\cite{Son:2009vu} for $\sig=0$.  Consider for example a free bulk scalar field $\Phi$. In Fourier space, with
\be
\Phi = \Phi (r, \om, \vk) e^{-i \om v + i \vk \cdot \vx},
\ee
a general solution can be written close to the horizon $r_h$ as
\be \label{unk}
\Phi (r,\omega,\vk) =A (\omega,\vk) (r- r_h)^{\frac{i\beta \omega}{2\pi}}+B (\omega,\vk)
\ee
where $\beta$ is the inverse temperature, and $A, B$ are some functions. Now applying the prescription
of Fig.~\ref{fig:contour}, we find that the corresponding coefficients $A_{1,2}, B_{1,2}$ for $\Phi_{1,2}$ on the upper and lower segments of the contour of Fig.~\ref{fig:contour} should be related by
\be
A_2 = A_1 e^{ \beta \om}, \qquad B_2 = B_1
\ee
which is precisely the prescription of~\cite{Son:2009vu}. Note that in this frequency space approach one cannot
easily perform small $\om$ expansion as in~\eqref{unk}, when expanding $\om$, one encounters powers of $\log (r-r_h)$
which become singular at the horizon $r = r_h$.

\section{A scalar example} \label{sec:3}

We will now illustrate the new prescription of the previous section using a scalar two-point function.
We are interested in extracting its large-distance and long-time behavior using derivative expansion.

\subsection{Equilibrium}\label{sec:eq}

We first consider an equilibrium state with metric~\eqref{metric} for which functions $f$ and $\lam_{ij}$ are $x^\mu$-independent.
Consider a massless scalar field
\be S=-\frac 12 \int d^{d+1}x\sqrt{-g} g^{MN} \p_M\Phi \p_N \Phi \ ,\ee
where $M=0,1,\dots,d$. The equation of motion for $\Phi$ can be written as
\be\label{phie}
{1 \ov \sqrt{\lam}} \p_r \le( f \sqrt{\lam} \p_r \Phi \ri)
+ \p_0 \p_r \Phi + {1 \ov \sqrt{\lam}} \p_r ( \sqrt{\lam} \p_0 \Phi)
+ \lam^{ij} \p_i  \p_j \Phi = 0 \
\ee
with boundary conditions
\be
\Phi (r, x^\mu) \to \phi_1 (x), \quad r \to \infty_1 , \qquad \Phi  (r, x^\mu) \to \phi_2 (x), \quad r \to \infty_2 ,
\ee
where $\infty_1, \infty_2$ denote the two asymptotic infinities (see Fig.~\ref{fig:contour}) and where $\p_0$ denotes derivative with respect to $v$. We are interested in solving~(\ref{phie}) in the regime where the boundary sources $\phi_1(x),\phi_2(x)$ are slowly varying in $x^\mu$ compared to the characteristic scales of the system, which can be taken to be the inverse temperature $\beta=4\pi/f'(r_h)$. More explicitly, we assume
\be \label{dhj}
 {\beta \p_\mu \phi_1 \ov \phi_1} \sim {\beta \ov L}  \ll 1, \qquad  {\beta \p_\mu \phi_2 \ov \phi_2} \sim {\beta \ov L}  \ll 1 \
\ee
where $L$ is the typical time/length scale associated with the variations of the sources.
We assume that $\Phi(r,x)$ satisfies the same condition for any $r$, i.e.
\be
 {\beta \p_\mu \Phi (r, x^\mu) \ov \Phi} \ll 1, 
\ee
while the derivatives with respect to $r$ can be arbitrary. We will see that this is a consistent assumption, thanks to the fact that the integration contour Fig.~(\ref{fig:contour}) goes around the horizon $r_h$, thus avoiding the singular behavior of~(\ref{unk}) at $r -r_h$. We can then expand $\Phi$ in terms of boundary derivatives of $\phi_{1,2}$ as
\be
\Phi = \Phi^{(0)} + \Phi^{(1)} + \Phi^{(2)} + \cdots
\ee
and solve~(\ref{phie}) order by order in the number of derivatives.  The boundary conditions for various $\Phi^{(n)}$ are
\bega\label{dir1}
\Phi^{(0)} (r, x^\mu) \to \phi_1 (x), \quad r \to \infty_1 , \qquad \Phi^{(0)} (r, x^\mu) \to \phi_2 (x), \quad r \to \infty_2 ,\\
\Phi^{(n)} (r) \to 0, \quad r \to \infty_{1}, \qquad \Phi^{(n)} (r) \to 0, \quad \quad r \to \infty_2 , \qquad
n \geq 1 \ .
\end{gather}
For our discussion below it is convenient to introduce the following combinations
\be
\phi_r = \ha (\phi_1 + \phi_2) \qquad  \phi_a = \phi_1 - \phi_2 \ .
\ee

At zeroth order we have
\be\label{eq0}
{1 \ov \sqrt{\lam}} \p_r \le( f \sqrt{\lam} \p_r \Phi^{(0)} \ri) = 0,
\ee
whose solution which satisfies the Dirichlet boundary conditions (\ref{dir1}) is given by
\be \label{sol0}
\Phi^{(0)} = - {\phi_a \ov Q_0} b_1 (r) 
+ \phi_1
= \phi_r - {\phi_a \ov  Q_0} b_r (r)
\ee
with
\bega\label{bQ}
b_1 (r) = \int_{\infty_1}^r {dr \ov \sqrt{\lam} f}, \qquad
 b_2 (r) = \int_{\infty_2}^r {dr \ov \sqrt{\lam} f}, \qquad b_r (r) = \ha \le(b_1 + b_2 \ri),  \cr
Q_0 \equiv  b_1 (\infty_2) = -b_2 (\infty_1)  = b_1 - b_2 
= \oint_{r_h} {dr \ov \sqrt{\lam} f}
= - {1 \ov \sqrt{\lam_h}}  {i \beta \ov 2} \ .
\end{gather}
The contour of integrals in (\ref{bQ}) is along that of Fig. \ref{fig:contour}. The direction that the integration goes around $r=r_h$ is dictated by the limits of the integrals. For example, for $r$ lying on the upper segment one goes around $r_h$ clockwise
in the integral for $b_1 (r)$, while in $b_2 (r)$, when $r$ lies in the lower segment, one goes around $r_h$ counterclockwise.
In $Q_0$ the integral goes clockwise around $r_h$, which determines the sign of its value $-\frac{i\beta} 2$. Note that it is crucial that the integrand for $b_1 (r)$ has a logarithmic divergence at $r_h$, which makes it possible for $\Phi^{(0)} (r, x^\mu)$ to satisfy both the boundary conditions at $\infty_{1,2}$. If the integrand were regular at $r=r_h$, then we would be forced to have $\phi_1 = \phi_2$.

At $n$-th order, we find
\be\label{311}
{1 \ov \sqrt{\lam}} \p_r \le( f \sqrt{\lam} \p_r \Phi^{(n)} \ri) = s^{(n)}
\ee
where $s^{(n)}$ denotes terms which were already known from lower orders, e.g.
\bega\label{312}
s^{(1)} = - \p_0 \p_r \Phi^{(0)} - {1 \ov \sqrt{\lam}} \p_r \le(\sqrt{\lam} \p_0 \Phi^{(0)} \ri) \cr
s^{(2)} = - \p_0 \p_r \Phi^{(1)} - {1 \ov \sqrt{\lam}} \p_r \le(\sqrt{\lam} \p_0 \Phi^{(1)} \ri)-
\lam^{ij} \p_i \p_j  \Phi^{(0)} \ .
\end{gather}
Equation for $\Phi^{(n)}$ can be solved in general  as
\be
\Phi^{(n)} (r) = \int_{\infty_1} ^r dr' \, {1 \ov \sqrt{\lam} f} \le[\int_{\infty_1}^{r'} dr'' \sqrt{\lam} s^{(n)}
+ c_n \ri]
\ee
where $c_n$ is chosen so that $\Phi^{(n)} (r \to \infty_2) \to 0$. For $n=1$,
after some manipulations we find
\be \label{sol1}
\Phi^{(1)} = {\p_0 \phi_a \ov Q_0} \le(a_r (r) b_r (r) - {1 \ov 4} \sqrt{\lam_h} Q_0^2 \ri)
+ {\p_0 \phi_r }  (\sqrt{\lam_h} b_r (r) -  a_r (r)  )
\ee
where we have further introduced
\be\label{a1}
a_1 (r) = \int_{\infty_1}^r {dr \ov f},  \qquad
a_2 (r) = \int_{\infty_2}^r {dr \ov f}, \qquad
 a_r (r) = \ha (a_1  + a_2 ) , \qquad a_1  - a_2 = \sqrt{\lam_h} Q_0 \ .
\ee
One easily sees that this procedure generalizes to all derivative orders. Comparing~\eqref{sol0} and~\eqref{sol1} we note that in order for the derivative expansion to make sense we need to choose the radius $\ep$ of the circle around the horizon to be not too small. Explicitly, as $\ep\to 0$, the first term in (\ref{sol0}) diverges logarithmically
\be \label{1i}
\Phi^{(0)}(r=r_h+\ep)\to -i\frac{\phi_a}{2\pi}\log\ep\ ,\ee
while the first term in (\ref{sol1}) has a stronger divergence
\be \Phi^{(1)}(r=r_u+\ep)\to i\beta \frac{\p_0\phi_a}{8\pi^2}(\log\ep)^2\ ,\ee
so, in order to have $\Phi^{(1)}$ subleading to $\Phi^{(0)}$, we need
\be \label{3i}
\phi_a\gg \beta\p_0\phi_a |\log\ep|  , \quad {\rm i.e.} \quad  1 \gg \ep \gg e^{- {L \ov \beta}}
\ee
where we have used~\eqref{dhj}. In the strict derivative expansion limit $L \to \infty$, so $\ep$ can be taken to zero at the end.

We now proceed to evaluate the on-shell action for $\Phi$, which will yield the Schwinger-Keldysh generating functional of the boundary theory to quadratic order in sources.  Plugging (\ref{sol0}) and (\ref{sol1}) in the bulk action, we will find the on-shell action $W$ up to second order in derivatives. To proceed, we write the bulk action as
\be\label{act2}
S =  -\ha \int d^{d+1} x \sqrt{\lam} \, \le[f (\p_r \Phi)^2 + 2 \p_r \Phi \p_0 \Phi + \lam^{ij} \p_i \Phi \p_j \Phi  \ri]
\ee
The on-shell action at zeroth order in derivatives is
\be
W^{(0)} = -\ha \int d^{d+1} x \sqrt{\lam} f (\p_r \Phi^{(0)})^2 =  \ha \int d^d x \, {\phi_a^2 \ov Q_0}
=   {i \ov \beta} \int d^d x \,\sqrt{\lam_h}  \phi_a^2 \ .
\ee
At first order we have
\be\begin{split}
W^{(1)}=&   - \int d^{d+1} x \, \sqrt{\lam} [f\p_r\Phi^{(0)}\p_r\Phi^{(1)}+ \p_r \Phi^{(0)} \p_0 \Phi^{(0)}]\\
=& - \int d^{d+1} x \sqrt{\lam}  \p_r \Phi^{(0)} \p_0 \Phi^{(0)}
 = - \int d^d x \, \sqrt{\lam_h} \phi_a \p_0 \phi_r\end{split}
 \label{W1}
 \ee
where the first term in the second expression can be shown to vanish by performing integration by parts, using the fact that $\Phi^{(0)}$ is a solution of (\ref{eq0}), and that $\Phi^{(1)}$ vanishes on the boundaries $\infty_1,\infty_2$.

The second order action will turn out to contain integrals which are divergent due to the behavior of various functions as $r\to \infty$, which is expected from standard near-boundary behavior of the bulk gravity theory \cite{Skenderis:2002wp}. To cure these divergences, one evaluates the integrals up to some cut-off slice $r_\Lambda<\infty$, add a local counterterm action $S\to S+S_{\text{ct}}$ to (\ref{act2}) to compensate for the divergence, and take $r_\Lambda\to\infty$ at the end. One additional ingredient here is that we are working with two boundaries, which means that we need to add one counterterm action per boundary:
\be \label{sct}\begin{split}S_{\text{ct}}=&\frac 1{d-2} r_{\Lambda}^{d-2}\int d^d x\eta^{\mu\nu}\p_\mu\phi_1\p_\nu\phi_1-\frac 1{d-2} r_\Lambda^{d-2}\int d^d x\eta^{\mu\nu}\p_\mu\phi_2\p_\nu\phi_2\\
=&\frac 1{d-2} r_\Lambda^{d-2}\int d^d x\eta^{\mu\nu}\p_\mu\phi_a\p_\nu\phi_r\ ,\end{split}\ee
where the power of $r_\Lambda$ and the coefficient in the prefactor are determined in order to compensate the divergence from bulk integrals, and $\eta_{\mu\nu}$ is the boundary Minkowski metric. Note that there cannot be terms proportional to $\phi_a^2$ as counterterms should always have a factorized structure as the first line of (\ref{sct}). Doing similar steps as above and including the counterterm action (\ref{sct}), the second order on-shell action can be expressed as
\be \begin{split}
W^{(2)} = & -\frac 12 \int^{r_\Lambda} d^{d+1}x\sqrt\lambda[\p_0\Phi^{(1)}\p_r\Phi^{(0)}  +\p_r\Phi^{(1)}\p_0\Phi^{(0)} +\lambda^{ij}\p_i\Phi^{(0)}\p_j\Phi^{(0)}]\\
&
+\frac 1{2(d-2)} r_\Lambda^{d-2}\int d^d x\eta^{\mu\nu}\p_\mu\phi_a\p_\nu\phi_r\\
=& \int d^d x \,  \le\{ Q_{ra} \p_0 \phi_a \p_0 \phi_r + Q_{ra}^{ij} \p_i \phi_a \p_j \phi_r+\frac{i}{2\beta}Q_{aa} (\p_0 \phi_a)^2 + {i
\ov 2\beta} Q_{aa}^{ij}\p_i \phi_a\p_j\phi_a  \ri\}
\end{split}\label{act3}\ee
where in the first line, the superscript $r_\Lambda$ in the integration means that we integrate in going from $r_\Lambda$ of the upper segment to $r_\Lambda$ of the lower segment, following the contour of Fig. \ref{fig:contour}, and at the end of the evaluation $r_\Lambda$ should be taken to $\infty_2$ and $\infty_1$ in the upper and lower segment, respectively. The coefficients in (\ref{act3}) are
\begin{eqnarray}\label{intQ}
 Q_{ra}=&\displaystyle-\frac 1{Q_0} \int^{r_{\Lambda}} dr\frac{\sqrt\lambda  b_r}f-\frac {r_\Lambda^{d-2}}{d-2},\qquad
Q_{aa}&=-i\frac \beta{ Q_0^2}\int_{\infty_2}^{\infty_1}\frac{dr}f\left(\sqrt\lambda b_r^2+\frac 14 \sqrt{ \lambda_h}Q_0^2\right)\\
Q_{ra}^{ij}=&\displaystyle \frac 1{Q_0}\int^{r_\Lambda}dr\sqrt{\lambda}\lambda^{ij}b_r+\frac{r_\Lambda^{d-2}}{d-2},\qquad Q_{aa}^{ij}&=i\frac\beta{Q_0^2}\int_{\infty_2}^{\infty_1}dr\sqrt\lambda \lambda^{ij}b_r^2\ .\label{intQ1}
\end{eqnarray}

For the Schwarzschild metric with $d=4$, we find
\be Q_{ra}=\frac 12 r_h^2(\log 2-1),\quad Q_{ra}^{ij}=\frac 12 r_h^2\delta_{ij}\ .\ee
Putting together first order and second order retarded action gives the retarded propagator
\be G_R(\omega,q)= r_h^3 i \omega+ \frac 12 r_h^2(\log 2-1)\omega^2 +\frac 12 r_h^2 q^2\ ,\ee
which agrees with the retarded propagator of the tensor mode in \cite{Policastro:2002se}.

One can readily verify that $W$ satisfies the KMS symmetry $W[\tilde\phi_1,\tilde\phi_2]=W[\phi_1,\phi_2]$, where \cite{Glorioso:2018wxw}
\be \tilde\phi_1(x^\mu)=\phi_1(-x^0+i\theta,x^i),\qquad \tilde\phi_2(x^\mu)=\phi_2(-x^0-i(\beta_0-\theta),x^i)\ .\ee
Note that, at this derivative order, KMS invariance does not impose any constraint on $W^{(2)}$. One can indeed check that second derivative terms generated by $W^{(1)}[\tilde\phi_1,\tilde\phi_2]$ cancel with each other.

We conclude this subsection by noting that the term in $W$ proportional to $\phi_a^2$ arises from the small circle around the horizon connecting the two horizontal segments. This is consistent with the general expectation that noises arise from fluctuations around the horizon.


\subsection{Slowly-varying horizon}\label{sec:3b}

Let us now consider a state which evolves with time. We consider a spacetime whose horizon changes slowly in the boundary coordinates $x^\mu$. We shall give a proof of principle that the above discussion can be generalized to such spacetime by performing an explicit check at first derivative order.

For concreteness, we consider the AdS-Schwarzchild metric:
\be\label{ebb}
ds^2=-f dv^2+2drdv+r^2 dx_i^2,\qquad f=r^2\left(1-\frac{(r_h(x^\mu))^d}{r^d}\right)\ ,
\ee
where $r_h(x^\mu)$ is a slowly-varying function, and we treat derivatives $\p_\mu r_h$ with the same power counting as in ~\eqref{dhj}.  To first order, we only need to account for the correction to $W^{(1)}$ which is proportional to $\p_0 r_h(v)$, which we denote by $\delta W^{(1)}$. From (\ref{W1}) we find
\be \delta W^{(1)}=-\int d^{d+1}x \sqrt\lambda \p_r \Phi^{(0)}\delta(\p_0\Phi^{(0)})\ ,\ee
where $\delta(\p_0\Phi^{(0)})$ is the part of $\p_0\Phi^{(0)}$ which is proportional to first derivatives of $r_h$:
\be \delta(\p_0\Phi^{(0)})=\frac{d  r_h^{d-1}}{2\pi r^{d-2} f}i \phi_a \p_0 r_h\ .\ee
One then finds
\be \delta W^{(1)}=\frac i2 \int d^d x B(x^\mu)\p_0 r_h(x^\mu )\phi_a^2\ ,\ee
where
\be \label{Bx}B(x^\mu)=i\frac{ d^2 r_h^{2d-1}}{2\pi ^2}\int_{\infty_2}^{\infty_1} dr\frac 1{r^{d-2}f^2}=\frac{d-2}\pi r_h^{d-2}
\ .\ee

Altogether, the action at first order reads
\be W^{(0)}+W^{(1)}+\delta W^{(1)}=\int d^d xr_h^{d-1}\left(\frac 12(\phi_a\p_0\phi_r-\phi_r\p_0\phi_a)-\frac i{2}\left(\frac{d r_h}{4\pi}-\frac{d-2}{\pi r_h} \p_0 r_h\right)\phi_a^2\right)\ .
\ee

Note that the KMS symmetry still holds, where the local inverse temperature is
\be \beta=\left(\frac{d r_h}{4\pi}-\frac{d-2}{\pi r_h} \p_0 r_h\right)^{-1}\ ,\ee
i.e. the relation between temperature and horizon $r_h$ receives a derivative correction, which is because we are not in equilibrium. Note that we still have KMS symmetry because we assume local equilibrium. We expect that the analytic continuation continues to hold when the dependence of $r_h$ on boundary spacetime coordinates is not slow.


\section{Effective field theory of diffusion from gravity}\label{sec:vec}

The generating functional we found in Sec.~\ref{sec:3} for a scalar operator is local, in the sense that the corresponding $W$ can be written
as $W = \int d^d x \, F [\phi_{1}, \phi_2]$ where $F$ is a function of $\phi_{1,2}$ and their derivatives, and
has a well-defined derivative expansion. For a conserved $U(1)$ current $J^\mu$ or the stress tensor $T^{\mu \nu}$, the corresponding generating functional will not be local as there are gapless hydrodynamic modes associated with
these conserved quantities, the integrating out of which leads to nonlocal behavior. The goal of hydrodynamic effective
action is to isolate the effective dynamics of these modes.

In this section we shall derive from gravity the effective action for the diffusion mode associated with a conserved $U(1)$ current at finite temperature. The story is more complex than the discussion of last section. In this case one cannot solve the full bulk equations of motion, as that would amount to integrating over all modes, diffusion modes included. Instead one should integrate out the degrees of freedom except for the one corresponding to diffusion. Deriving effective actions (at non-dissipative level) of hydrodynamic modes from gravity was initiated by~\cite{Nickel:2010pr}, and subsequent work include~\cite{Crossley:2015tka,deBoer:2015ija}.

Below we first present a quick review of the formulation of effective field theory for a diffusion mode given in~\cite{CGL},
in anticipation of the structures that we will find in the gravity derivation.

\subsection{Review of EFT for diffusion}

Consider a quantum system with a conserved $U(1)$ current $J^\mu$ at a finite inverse temperature $\beta$. The corresponding Schwinger-Keldysh generating functional is $W [A_{1\mu} , A_{2 \mu}]$, where $A_{1\mu}$ and $A_{2\mu}$ are
sources for $J^\mu$ along two segments of the contour,
\be\label{gent0}
e^{i W [A_{1\mu} , A_{2 \mu}]} \equiv  
\Tr \left[\rho_0 \sP e^{ i \int  d^d x  \, (A_{1\mu} J_1^\mu -A_{2\mu} J_2^\mu) }\ri]\ .
\ee
Due to current conservation, the generating functional satisfies
\be \label{Wg} W[A_{1\mu},A_{2\mu}]=W[A_{1\mu}+\p_\mu \lambda_1,A_{2\mu}+\p_\mu \lambda_2]\ ,
\ee
where $\lambda_1,\lambda_2$ are independent functions. The generating functional $W [A_{1\mu} , A_{2 \mu}]$
can be obtained from the path integrals over a pair of diffusion fields $\vp_{1,2}$
\be \label{ieft}
e^{i W [A_{1\mu} , A_{2 \mu}]} =\int D\varphi_1 D\varphi_2 \, e^{iI_{\text{EFT}}[B_{1\mu},B_{2\mu}]}\ ,\ee
where
\be \label{sty}
B_{1\mu}=A_{1\mu}+\p_\mu\varphi_1,\qquad B_{2\mu}=A_{2\mu}+\p_\mu\varphi_2\ ,
\ee
and $I_{\text{EFT}}[B_{1\mu},B_{2\mu}]$ is a local action of $B_1, B_2$. The combinations~\eqref{sty} ensure that
(\ref{Wg}) is satisfied after integrating out $\varphi_1,\varphi_2$, and furthermore, equations of motion for $\vp_{1,2}$ are
equivalent to imposing the current conservation along two segments of the contour.
$I_{\text{EFT}}$ can be obtained as the most general local action of $B_1, B_2$ which are: (i) translationally invariant; (ii) rotationally invariant; (iii) invariant under the following diagonal shift symmetry\footnote{The origin of (\ref{ds}) is due to that the $U(1)$ symmetry associated with current conservation is not spontaneously broken. We then have the freedom of independently relabeling the phase of each charged constituent of the system on one given time slice.
As we are considering a global symmetry, the phase redefinition cannot depend on time $x^0$, but since the constituents of the system are independent of one another,  they should have the freedom of making independent phase rotations, i.e. we should allow phase rotations of the form
$e^{i \lam (x^i)}$
. 
When the $U(1)$ symmetry is spontaneously broken, i.e. the system is in a superfluid phase, the phases of all the constituents are locked together, and (\ref{ds}) should be dropped.
}
\be\label{ds}
\varphi_1 (x^0, x^i) \to  \varphi_1 (x^0, x^i)  + \lam (x^i) , \qquad \varphi_2 (x^0, x^i) \to  \varphi_2 (x^0,  x^i)  + \lam (x^i)
 \ ,
\ee
(iv) invariant under dynamical KMS symmetry. With $I_{\text{EFT}}=\int d^dx \mathcal L$, then to second order in derivatives and quadratic in $B_{1,2}$,  $\sL$ can be written as as
\begin{gather}
 \sL  = {i \ov 2} a_{00} B_{a0}^2 + {i \ov 2} b_{00} B_{ai}^2 +  g_{00} B_{a0} B_{r0}
+ i f_{00} B_{a0} \p_i B_{ai}  + g_{10} \p_0 B_{a0}  B_{r0}  \cr + u_{00} \p_i B_{ai} B_{r0}   +
 v_{00}  B_{ai} \p_0 B_{ri}
+ {i \ov 2} a_{20} (\p_0B_{a0})^2 + {i \ov 2} a_{02} (\p_i B_{a0})^2  \cr + i f_{10} \p_0 B_{a0} \p_i B_{ai}
 + {i \ov 2} b_{20} (\p_0 B_{ai})^2  +  {i \ov 2} b_{02} (\p_i B_{aj})^2 + {i \ov 2} c_{00} (\p_i B_{ai})^2
+
g_{20} \p_0 B_{a0} \p_0 B_{r0} \cr
+ g_{02} \p_i B_{a0} \p_i B_{r0} + h_{00}  B_{a0} \p_i \p_0 B_{ri}
   + v_{10} \p_0 B_{ai} \p_0 B_{ri}
   - u_{10} \p_i B_{ai} \p_0 B_{r0}+\frac{w_{00}}2F_{aij}F_{rij}
   \ ,
   \label{ac2n}
 \end{gather}
\newpage
\hspace{-0.5cm}where $a_{00}, b_{00}, \cdots$ are constants and we have introduced
\be\begin{gathered}
B_{r \mu} = \ha (B_{1 \mu} + B_{2 \mu}), \qquad B_{a \mu} =B_{1 \mu} - B_{2 \mu}\\
F_{rij}=\p_i B_{rj}-\p_j B_{ri},\qquad F_{aij}=\p_i B_{aj}-\p_j B_{ai}
 \ ,
\end{gathered}\ee
The explicit expressions of the currents can be found by varying $I_{\text{EFT}}$ with respect to $B_{r\mu}$ and $B_{a\mu}$. Introducing
\be
J_r^\mu=\frac 12(J_1^\mu+J_2^\mu), \qquad J_a^\mu=J_1^\mu-J_2^\mu,
\ee
 one finds
\be\label{Ja} J_r^\mu=\frac{\delta I_{\text{EFT}}}{\delta B_{a\mu}},\qquad
 J_a^\mu=\frac{\delta I_{\text{EFT}}}{\delta B_{r\mu}}\ .\ee
More details, elaborations and generalizations of the above action can be found in~\cite{CGL}.

\subsection{Obtaining generating functional $W[A_1, A_2]$} \label{sec:genb}

Before discussing how to obtain the effective action $I_{\rm EFT} [B_1, B_2]$ for diffusion fields $\vp_{1,2}$, we first
discuss how to obtain the full generating functional $W[A_1, A_2]$ of~\eqref{gent0}, in a manner which helps motivating the
prescription for obtaining $I_{\rm EFT} [B_1, B_2]$.

Suppose $J^\mu$ is dual to a bulk gauge field $C_M$ with an action
\be\label{actp} S=-\frac 14 \int d^{d+1}x \sqrt{-g} F_{MN}F^{MN}\ ,\ee
where $F_{MN}=\p_MC_N-\p_N  C_M$.
The background spacetime is again~(\ref{metric}) with $f, \lam_{ij}$ being $x^\mu$ independent.
 The boundary conditions are
\be\label{hek}
C_\mu (r \to \infty_1) = A_{1\mu}, \qquad C_\mu (r \to \infty_2) = A_{2\mu}
\ee
with $A_{1\mu}, A_{2 \mu}$ as sources for $J_1^\mu, J_2^\mu$. Equations of motion of~(\ref{actp}) are
\be\label{ujn}
E^M\equiv \frac 1{\sqrt{-g}}\p_N(\sqrt{-g}F^{MN})=0\ .
\ee
It is also convenient to introduce
\be \label{can}
\Pi^\mu=\sqrt{-g}F^{r\mu}
\ee
which is the conjugate momentum to $C_\mu$, with ratial direction $r$ treated as ``time.''
Recall that when evaluated on the solutions to equations of motion, the boundary values of $\Pi^\mu$
simply give the expectation values of boundary currents $J^\mu_1$ and $J^\mu_2$.  The radial component of the bulk equations (\ref{ujn}) can be written in terms of~\eqref{can} as
\be E^r=\frac 1{\sqrt{-g}}\p_\mu\Pi^\mu=0\ ,\label{conse}\ee
which are then equivalent to the conservation of $J^\mu_{1,2}$ when evaluated at the boundary.
From the Bianchi identity, which can be written as
\be
 \p_r (\sqrt{-g}E^r) + \p_0 (\sqrt{-g}E^0) + \p_i (\sqrt{-g}E^i) = 0 ,
\ee
it is enough to impose $E^r =0$ at a single slice of $r$.

Now consider performing a  gauge transformation
\be \label{fixr}C_M (r,x^\mu)\to C_M (r,x^\mu)+\p_M \Lambda(r,x^\mu),\qquad \Lambda(r,x^\mu)=\int_r^{r_c} dr' C_r(r',x^\mu)\
\ee
to set $C_r$ to zero. Here $r_c$ is an arbitrary reference point, which for later convenience we will take
to lie on the small circle around $r_h$. The boundary conditions~\eqref{hek} are modified by such a gauge transformation to
\be \label{onn}
C_\mu (r \to \infty_1) =  B_{1\mu} = A_{1\mu} + \p_\mu \varphi_1 , \qquad C_\mu (r \to \infty_2) = B_{2\mu} = A_{2 \mu}  + \p_\mu \varphi_2 
\ee
where
\be \label{onn1}
\varphi_1 = \int_{\infty_1}^{r_c} dr \, C_r , \qquad \varphi_2 = \int_{\infty_2}^{r_c} dr \, C_r \ .
\ee
As in previous sections all the $r$-integrals should be viewed as contour integrals along the complex $r$-contour
of Fig.~\ref{fig:contour} with directions of integrations specified by the integration limits.

Note that $\vp_{1,2}$ should be considered as dynamical variables: they are the Wilson line degrees of freedom associated
with $C_r$. They are not independent, as in the gauge $C_r =0$ there is still a residual gauge transformation $C_\mu \to C_\mu + \p_\mu \lam (x^\mu)$ with $\lam (x^\mu)$ an arbitrary function of boundary coordinates. Under this transformation, $\varphi_1\to \varphi_1+\lambda$  and $\varphi_2\to \varphi_2+\lambda$, thus we could shift one of them away, and the combination $ \varphi_1 - \varphi_2$ remains.

The generating functional $W[A_1, A_2]$ of~\eqref{gent0} can be obtained from a two-step procedure:

\ben

\item Find the gauge fixed on-shell action $\tilde S_{\rm on-shell} [B_1 , B_2] $ by plugging solution $C_\mu$ into equations of motion $E^\mu =0$ with boundary conditions~\eqref{onn}. Note that one does not impose equation of motion $E^r =0$ for $C_r$.
From the above discussion, $\tilde S$ is a functional of $A_1, A_2$ and $\vp_1 - \vp_2$.

\item Extremizing $\tilde S_{\rm on-shell}$ with respect to one of $\vp$'s, say $\vp_1$,
\be \label{omo}
{\de \tilde S_{\rm on-shell} \ov \de \vp_1} = 0,
\ee
which corresponds to imposing $E^r =0$ at
slice $r =\infty_1$. Then $W[A_1, A_2]$ is equal to $\tilde S_{\rm on-shell}$ evaluated on the solution $\vp_1$.

\een

\subsection{Prescription for finding $I_{\rm EFT} [B_1, B_2]$}

\begin{figure}[!h]
\begin{center}
\includegraphics[width=12.5cm]{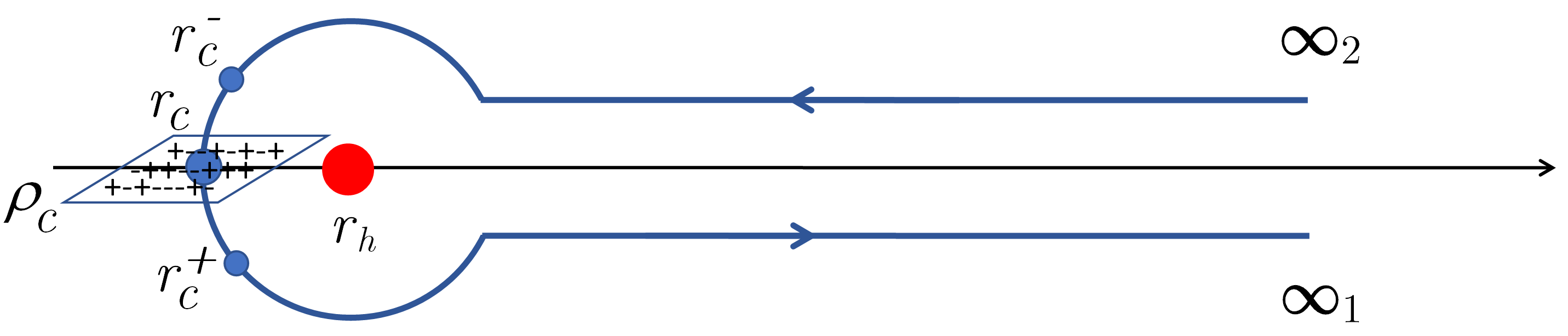} \end{center}
\caption{
A boundary condition~\eqref{hbd0} is imposed at the ``stretched horizon" $r=r_c$. As will be discussed below, such a boundary condition implies presence of a nonzero surface  charge $\rho_c$ at $r=r_c$, with a discontinuity in the radial component of the electric field
at $r_c$.}
 \label{fig:contour3}
\end{figure}

Comparing~\eqref{onn} with~\eqref{sty} we would like to identify~\eqref{onn1} as the hydrodynamic fields for diffusion.
Clearly to derive the effective action $I_{\rm EFT} [B_1, B_2]$,  we need to skip the step~\eqref{omo}, as $\vp_{1,2}$ equations of motion correspond to imposing $\p_\mu J^\mu_{1,2} =0$ at the boundary.
 But this is not enough,
as discussed above the gauge fixed on-shell action $\tilde S_{\rm on-shell} [B_1 , B_2] $ depends only on the combination $\vp_1 - \vp_2$, while we want independent $\vp_{1,2}$. For this purpose we will impose an additional boundary condition
\be \label{hbd0}
C_0 (r_c) = 0
\ee
where recall that $r_c$ is the reference point used to define $\vp_{1,2}$ in~\eqref{onn1}, see Fig.~\ref{fig:contour3}.
$r_c$ may be viewed as  a ``stretched horizon'' and its precise location on the small circle is not important.
Note that due to the residual gauge symmetry from fixing $C_r =0$, one could equivalently choose any fixed function of $x^\mu$
to be on the right hand side of~\eqref{hbd0}.

After imposing~\eqref{hbd0}  there is still a residual gauge freedom $C_i \to C_i + \p_i \lam (x^i)$ where the gauge transformation $\lam (x^i)$ depends only  on the boundary spatial coordinates. Such a gauge transformation also shifts the boundary values of $C_{\mu}$ leading to the shifts
\be\label{cs}
\varphi_1 (x^0, x^i) \to  \varphi_1 (x^0, x^i)  + \lam (x^i) , \qquad \varphi_2 (x^0, x^i) \to  \varphi_2 (x^0,  x^i)  + \lam (x^i)
 \ .
\ee
This precisely corresponds to the diagonal shift symmetry~\eqref{ds} for the EFT. 
Here we see the ``dual'' statement: on the gravity side it is a consequence of a boundary condition at the horizon.

To summarize, our prescription for computing the effective action in~\eqref{ieft} for diffusion is
\be\label{pref}
I_{\rm EFT} [B_{1 \mu}, B_{2 \mu}]  = \tilde S_{\rm on-sell} |_{C_0 (r_c) = 0}
\ee
where the right hand side means that the solution to $E^\mu =0$ should satisfy both boundary conditions~\eqref{onn} and~\eqref{hbd0}. In fact, as we will see in next subsection when analyzing the explicit equations, the independent boundary conditions for $C_0$ at $\infty_1$ and $\infty_2$ are only possible with the imposing of~\eqref{hbd0}.

Now let us elaborate a bit further on the physical
meaning and implications of~\eqref{hbd0}. Imposing~\eqref{hbd0} means that the equation of motion for $C_0$ at $r_c$
is not imposed as we are not free to vary the value $C_0$ at that point. More explicitly, the $0$-th component of~\eqref{ujn} can be written as
\be
 - \p_r \Pi^0 + \p_i \le(\sqrt{-g} F^{0i} \ri) = 0  \ .
\ee
For any $r$, integrating the above equation from $r_c^- = r_c- \de$ to $r_c^+ = r + \de$ with $\de \to 0$ we find that
\be
\Pi^0 (r^+_c) - \Pi^0 (r_c^-) = 0 \ .
\ee
Thus not imposing $E^0 = 0$ at $r=r_c$ means that there can be discontinuity in $\Pi^0$ at $r_c$, i.e. in general
\be \label{rhoc}
\rho_c \equiv \Pi^0 (r^+_c) - \Pi^0 (r^-_c)  \neq 0  \ .
\ee
Since $\Pi^0 = \sqrt{-g} F^{r0}$ is the radial component of the electric field, the above equation can be interpreted that there is a surface charge density proportional to $\rho_c$ at the hypersurface $r = r_c$, see Fig.~\ref{fig:contour3}.

Equation~\eqref{pref} can be written more explicitly as
\be
I_{\rm EFT} [B_{1 \mu}, B_{2 \mu}]  = \tilde S_1 [C_\mu (r_c), B_{1 \mu}] + \tilde S_2 [C_\mu (r_c), B_{2 \mu}]
\ee
where $\tilde S_1$ denotes the part of the on-shell action evaluated  for the region $(r_c^+, \infty_1)$ while
$\tilde S_2$ is that for the region  $(r_c^-, \infty_2)$. Now consider the variation of $\tilde S_1$ under a gauge transformation
$C_\mu \to C_\mu + \p_\mu \lam (x^\mu)$, we have
\be
0 = \int d^d x \,\le[ {\de S_1 \ov \de B_{1 \mu}} \p_\mu \lam + {\de S_1 \ov \de C_\mu (r_c)} \p_\mu \lam \ri]
\ee
which leads to
\be \label{p1}
\p_\mu \Pi^\mu (\infty_1) = \p_\mu \Pi^\mu (r_c^+)   \ .
\ee
Similarly from $\tilde S_2$ we have
\be \label{p2}
\p_\mu \Pi^\mu (\infty_2) = \p_\mu \Pi^\mu (r_c^-)  \ .
\ee
Taking the sum and difference of~\eqref{p1}--\eqref{p2} we find that
\be \label{p3}
\p_\mu J_a^\mu =  \p_0 \rho_c+\p_i(\Pi^i(r_c^+)- \Pi^i(r_c^-)), \qquad \p_\mu J_r^\mu =  \ha \p_\mu \le( \Pi^\mu (r_c^+)  +  \Pi^\mu (r_c^-) \ri) \ .
\ee

Finally, let us note that one possible concern with the boundary condition~\eqref{hbd0} is whether it can be consistently imposed as $C_0$ does couple to other components of the gauge field. Discontinuity of $\Pi^0$ can potentially lead to discontinuities in $\Pi^i$ and make the other components of equations of motion not self-consistent. We will pay particular attention to these aspects in our discussion below and show it does not happen (more details in Sec.~\ref{sec:nhb}).

\subsection{Explicit bulk solution}\label{sec:bulks}

In this subsection we solve explicitly the bulk equations of motion $E^\mu =0$ with boundary conditions~\eqref{onn} and~\eqref{hbd0}
 order by order in derivatives, assuming the sources are slowly varying.  We discuss the general structure of the derivative expansion of the equations of motion and solve them up to first derivative order. For simplicity from now on we will take $\lambda_{ij}$ in  (\ref{metric}) to be diagonal.

The equations of motion $E^\mu =0$ can be written explicitly as
\bea \label{pi0e}
\p_r \Pi^0 &= & - \p_i \le(\sqrt{\lambda} \lambda^{ii} \p_r C_i\ri)
 \\
\p_r \Pi^i & = &  \p_0 \le(\sqrt{\lambda} \lambda^{ii} \p_r C_i\ri) +\sqrt{\lambda} (\lambda^{ii})^2 \sum_{j } \p_j  F_{ji}  \
\label{piie}
\eea
with
\be\label{conj}
\Pi^0 =  \sqrt{\lambda} \p_r C_0,\qquad \Pi^i =   - \sqrt{\lambda} \lambda^{ii} (f \p_r C_i + \p_0 C_i - \p_i C_0)   \ .
\ee
Like in the scalar case, we solve (\ref{pi0e})--(\ref{piie}) by doing derivative expansion of $C_\mu$,
\be
C_\mu= C_{\mu}^{(0)} + C_{\mu}^{(1)}  + C_{\mu}^{(2)}  + \cdots \ .
\ee
The boundary conditions at infinities are
\begin{eqnarray}\label{dir2}
C_\mu^{(0)} (r, x^\mu) &\to& B_{1\mu} (x), \quad r \to \infty_1 , \qquad C_\mu^{(0)} (r, x^\mu) \to B_{2\mu} (x), \quad r \to \infty_2 ,\\
C_\mu^{(n)} (r) &\to& 0, \quad r \to \infty_{1}, \qquad C_\mu^{(n)} (r) \to 0, \quad \quad r \to \infty_2 , \qquad
n \geq 1 \ .
\end{eqnarray}

At zeroth order, the equations for $C_0^{(0)}$ and $C_i^{(i)}$ read
\be\label{0eq}
\p_r \le(\sqrt{\lambda} \p_r C_0^{(0)} \ri) = 0, \qquad \p_r \le(\sqrt{\lambda} \lambda^{ii} f \p_r C_i^{(0)}\ri) = 0 \ .
\ee
The solution for $C^{(0)}_0$ can be written as
\be
C_0^{(0)} = \bca
k_{1} b_0(r)+h_{1}  & r_c\leq r\leq\infty_1 \cr
k_2 b_0(r)+h_2   & r_c\leq r\leq\infty_2
\eca \ ,
\ee
where $k_1,k_2,h_1,h_2$ are arbitrary functions of $v$, and
\be\label{1wq}
 b_0(r)=\int_{r_c}^r\frac{dr}{\sqrt{\lambda}},\qquad Q_0=b_0(\infty)\ .\ee
Imposing the UV boundary conditions (\ref{dir2}), together with the boundary condition at the horizon (\ref{hbd0}), we find the unique zeroth order solution
\be \label{a00}
C_0^{(0)} = \bca
{ B_{10} \ov Q_0} b_0 (r)  & r_c\leq r\leq\infty_1 \cr
{ B_{20} \ov Q_0} b_0 (r)  & r_c\leq r\leq\infty_2
\eca \ .
\ee
The equation for $C_i$ is almost identical to that of the scalar, eq. (\ref{eq0}), and can be solved all the way from $\infty_1$ to $\infty_2$
\be\label{ai0}
C_i^{(0)} 
=  B_{ri} - { B_{ai} \ov  Q_i} b_{ri} (r)  \ ,
\ee
where we have already imposed the boundary conditions (\ref{dir2}), $Q_0$ was defined in~(\ref{bQ}), and
\be b_{ri}(r)=\int_{\infty_1}^r dr\frac{\lambda_{ii}}{\sqrt \lambda f}-\frac 12 Q_i,\qquad Q_i=\int_{\infty_1}^{\infty_2} dr\frac{\lambda_{ii}}{\sqrt \lambda f}=-\frac 1{K_h}\frac{i\beta}2,\quad K_h=\left.\frac{\sqrt\lambda}{\lambda_{ii}}\right|_{r_h}\ .\ee

Notice from~\eqref{0eq} that the equation for $C_0$ is regular across the horizon, while the equation for $C_i$ has a similar structure to~\eqref{eq0} of the scalar. The difference is reflected in that the integrand of $b_0$ in~\eqref{1wq} is completely regular at $r=r_h$.  Without the boundary condition~\eqref{hbd0}, then we would not be able to satisfy independent boundary conditions at $\infty_1$ and $\infty_2$.

At $n$-th order the equations for $C_0^{(n)}$ and $C_i^{(n)}$ read
\be\label{eomn}
{1 \ov \sqrt{\lambda}} \p_r \le(\sqrt{\lambda} \p_r C_0^{(n)} \ri) = s_0^{(n)}, \qquad {1 \ov \sqrt{\lambda}}  \p_r \le(\sqrt{\lambda} \lambda^{ii} f \p_r C_i^{(n)}\ri) = s_i^{(n)}
\ee
where the right hand sides are expressed solely in terms of $(n-1)$th-order fields:
 \be\begin{split}
 s_0^{(n)} =& - \lambda^{ii} \p_i \p_r C_i^{(n-1)}\\
  s_i^{(n)} =& - \lambda^{ii} \p_0 \p_r C_i^{(n-1)} - {1 \ov \sqrt{\lambda}}  \p_r \le(\sqrt{\lambda} \lambda^{ii} (\p_0 C_i^{(n-1)} - \p_i C_0^{(n-1)})\ri) \\
 &  - \lambda^{ii} \lambda^{jj} \p_j \left(\p_j C_i^{(n-2)} - \p_i C_j^{(n-2)}\right)\ .
 \end{split} \ee
We now solve (\ref{eomn}) to find $C_0^{(1)},C_i^{(1)}$. Using (\ref{a00}) and (\ref{ai0}), at first order we have
\be
 s_0^{(1)} =  {\p_i B_{ia} \ov Q_i} {1 \ov \sqrt{\lambda} f}, \qquad s_i^{(1)} = {\p_0 B_{ia} \ov Q_i} {1 \ov \sqrt{\lambda} f} - {1 \ov \sqrt{\lambda}}  \p_r \le(\sqrt{\lambda} \lambda^{ii} (\p_0 C_i^{(0)} - \p_i C_0^{(0)})\ri) \ .
 \ee
We find the temporal component $C_0^{(1)}$ repeating the steps done to find (\ref{a00}), yielding
\be \label{a01}
C_0^{(1)} = {\p_i  B_{ia} \ov Q_i} \int_{r_h}^r {dr' \ov \sqrt{\lambda}} \le[\int_\infty^{r'} {dr'' \ov f} - {\tilde Q_0 \ov Q_0} \ri], \qquad \tilde Q_0 = \int_{r_h}^\infty {dr' \ov \sqrt{\lambda}} \int_\infty^{r'} {dr'' \ov f}
\ee
for both region I and region II. The spatial components $C_i^{(1)}$ give
\be \label{ai1}
C_i^{(1)} = - a_1 (r) \p_0 C_i^{(0)} (r) 
+ \int_{\infty_1}^r {dr \ov f} \p_i C_0^{(0)} + c_1 b_{1i} (r)\ ,
\ee
where for compactness we wrote the right hand side in terms of derivatives of $C_0^{(0)},C_i^{(0)}$, $a_1(r)$ is defined in eq. (\ref{a1}), and $c_1$ is an integration constant to  ensure $C_i^{(1)} (\infty_2) =0$,
\be
c_1 = K_h \p_0  B_{i2} + {\p_i  B_{0a} \ov Q_0 Q_i} \int_{r_h}^\infty {dr \ov f} b_0 (r) \ .
\ee
This concludes the evaluation of the bulk solution up to first order.

As anticipated, the conjugate momentum $\Pi^0$ of $C_0$ is discontinous at the horizon. We find that at the horizon
\be \label{disp}
\Pi^0(r_c^+) - \Pi^0(r_c^-) = { B_{0a} \ov Q_0} + K_h {\p_i  B_{ia} }\ .
\ee
Additionally, the non-analiticity in $C_0$ induces a non-analiticity in $C_i$. More explicitly, we can write
\be\label{vec2}
 C_i^{(1)} = \le( C_i^{(1)}  \ri)_{ana} +  \le( C_i^{(1)}  \ri)_{non}
 \ee
 where
 \be
 \le( C_i^{(1)}  \ri)_{ana}  = - a_1 (r) \p_0 C_i^{(0)} (r) + c_1 b_{i1} (r), \qquad
  \le( C_i^{(1)}  \ri)_{non} =  \int_{\infty_1}^r {dr \ov f} \p_i C_0^{(0)}
 \ee
and the subscript $ana$ and $non$ denote the analytic and non-analytic part respectively.

As in the discussion of~\eqref{1i}--\eqref{3i}, for the perturbation series to be valid near $r_h$ we need $\ep$ to lie in the range of~\eqref{3i}.

\subsection{Near-horizon behavior to all orders}\label{sec:nhb}
Below we shall analyze the near-horizon behavior of $C_0$ and $C_i$ to all derivative orders, which is a crucial test of consistency of our prescription.


To proceed, we first divide $C_i$ into analytic and non-analytic part
\be
 C_i = \le( C_i  \ri)_{\rm ana} +  \le( C_i  \ri)_{\rm non} \ .
 \ee
The analytic part $\le( C_i  \ri)_{\rm ana}$ does not have a well-defined value at the horizon and is defined
via analytic continuation around the horizon through an infinitesimal contour.
$C_0$ and the non-analytic part $\le( C_i  \ri)_{\rm non}$ are finite  at the horizon and are continuous, but
their derivatives are not. Similarly we can write $\Pi^i$ as
\be
 \Pi^i = \le( \Pi^i  \ri)_{\rm ana} +  \le( \Pi^i  \ri)_{\rm non} \ .
 \ee

Now let us look at the qualitative behavior to all orders for various quantities.
From~\eqref{eomn} we find that the equations of motion for the analytic part of $C_i^{(n)}$ can be written
schematically as (we will suppress the subscript ``ana'' for notational simplicity)
\be
\p_r (f \p_r C_i^{(n)}) = \p_r \p_0 C_i^{(n-1)} +  \p_0 C_i^{(n-1)}  + \p_i \p_j C_i^{(n-2)}
\ee
we thus find that near the horizon
\be\label{beha}
\le( C_i^{(n)}  \ri)_{\rm ana} \sim \le(\log (r-r_h) \ri)^{n+1} , \qquad
\le( \Pi^i_{(n)}  \ri)_{\rm ana} \sim  \le(\log (r-r_h) \ri)^{n} \ .
\ee
Now let us look at the non-analytic part. From~\eqref{eomn} we find that
\be \label{non1}
C_0^{(n)} \sim (r-r_h) (\log (r-r_h))^n, \qquad  \le( C_i^{(n)}  \ri)_{\rm non}  \sim (r-r_h) (\log (r-r_h))^{n-1} \ .
\ee
For $\Pi^\mu$ we find, using (\ref{conj}),
\be \label{non2}
\Pi^0_{(n)} \sim (\log (r-r_h))^n, \qquad  \le( \Pi^i_{(n)}  \ri)_{\rm non} \sim (r-r_h)  \le(\log (r-r_h) \ri)^{n-2} \ .
\ee
Note that the exponent for $\log (r-r_h)$ is one higher in~\eqref{non1} for $C_i$ than that for $\Pi^i$ in~\eqref{non2} as in $\Pi^i$ the leading non-analytic behavior from $C_i$ cancels with that from $C_0$.
The non-analytic part for $\Pi^i$ should appear at second order and it goes to zero on the horizon $r=r_h$.

The first equation of~\eqref{p3} implies that when we impose conservation of $J_a^\mu$, the divergence of the discontinuous part of the momentum at the horizon vanishes, i.e. $\p_\mu\Pi^\mu(r_c^+)-\p_\mu \Pi^\mu(r_c^-)=0$.
Eq. (\ref{non2}) shows that $\Pi^i(r_c^+)- \Pi^i(r_c^-)\to 0$ as $\ep \to 0$. In this limit, the first eq. in (\ref{p3}) then becomes
\be \p_\mu J^\mu_a=\p_0\rho_c\ .\ee
After imposing conservation of $J_a^\mu$, the surface charge $\rho_c$ located on the horizon becomes then time-independent. In particular, with the initial condition at $v = -\infty$ that $\rho_c =0$, we find that $\rho_c =0$ for all times, recovering the equation of motion for $C_0$ at $r_c$.


\subsection{The effective action}\label{sec:act}

Plugging the solution of section \ref{sec:bulks} into the bulk action
\be
S = - \ha \int d^{d+1} x \, \sqrt{-g} \le[f g^{ii} F_{ri}^2 + 2 g^{ii} F_{0i} F_{ri} - F_{0r}^2
+ \ha F_{ij} F^{ij} \ri]\ ,
\ee
evaluating the radial integral and taking $\ep \to 0$, one obtains the effective action for hydrodynamic modes $\vp_{1,2}$. We find that~\eqref{ac2n} is precisely recovered. The zeroth and first order coefficients are easily evaluated:
\be\label{459} b_{00}=\frac 1{i Q_i},\quad g_{00}=\frac 1{Q_0},\quad f_{00}=\frac{i}{Q_i Q_0}\int_{r_h}^{\infty}dr\frac{b_0}f,\quad v_{00}=-K_h,\quad a_{00}=g_{10}=u_{00}=0\ .\ee
The second order coefficients are
\bega
a_{20}=0, \qquad
g_{20}=0, \qquad
b_{20}=-\frac{\beta K_h}{8}-\frac i{Q_i^2}\int_{\infty_1}^{\infty^2} \frac{\sqrt{-g} b_{ri}^2}{\lambda_{ii}f},\\
v_{10}=\frac1{Q_i}\int_{\infty_1}^{\infty_2}\frac{a_1}f-\frac{\beta K_h}{4i}-\frac 1{Q_i}\int_{\infty_1}^{\infty_2}\frac{\sqrt{-g}b_{ri}}{\lambda_{ii}f},\\
f_{10}=-\frac i{Q_iQ_0}\int_\infty^{r_h}\frac{\sqrt{-g}b_{ri}b_0}{\lambda_{ii}f}-\frac i{Q_i Q_0}\int_\infty^{r_h}\frac{b_0}f\int_\infty^r\frac 1f-\frac i2\frac1{Q_0}\int_\infty^{r_h}\frac{\sqrt{-g}b_0}{\lambda_{ii}f},\\
h_{00}=-\frac{K_h}{Q_0}\int_\infty^{r_h}\frac{b_0}f+\frac 1{Q_0}\int_{\infty}^{r_h}\frac{\sqrt{-g}b_0}{\lambda_{ii}f},\qquad
a_{02}=0,\qquad
g_{02}=\frac 1{Q_0^2}\int_\infty^{r_h}\frac{\sqrt{-g}b_0^2}{\lambda_{ii}f},\\
u_{10}=\frac{\beta}{2i Q_i Q_0}\int_\infty^{r_h}\frac{b_0}f+\frac1{Q_0}\int_\infty^{r_h} \frac{\sqrt{-g}b_0}{\lambda_{ii}f},
\end{gather}
\bega
c_{00}=-\frac{2i}{Q_i^2}\int_{\infty_1}^{\infty_2}\frac 1f\int_{r_h}^r\frac{dr'}{\sqrt{-g}}\left(\int_\infty^{r'}\frac{dr''}f-\frac{\tilde Q_0}{Q_0}\right)-\frac{i}{Q_i^2}\int_{\infty_1}^{\infty_2}\frac 1{\sqrt{-g}}\left(\int_\infty^r\frac {dr'}f-\frac{\tilde Q_0}{Q_0}\right)^2,\\
b_{02}=\frac i{Q_i^2}\int_{\infty_1}^{\infty_2}\sqrt{-g}(\lambda^{ii})^2b_{ri}^2,\qquad
w_{00}=\frac 1{Q_i}\int_{\infty_1}^{\infty_2}\sqrt{-g}(\lambda^{ii})^2b_{ri}\ .
\end{gather}

Due to the asymptotic behavior of metric and gauge field at infinity, some of the integrals will be divergent for the same reason explained above (\ref{sct}). We again add a counterterm $S\to S+S_{\text{ct}}$, where
\be S_{\text{ct}}=\frac{C_d}4\int d^d x (F_1^{\mu\nu}F_{1\mu\nu}-F_2^{\mu\nu}F_{2\mu\nu})\ ,\ee
with $C_d= \frac 1{(d-4)}r_\Lambda^{d-4}$ for $d\neq 4$, and $C_4=\log\frac{r_\Lambda}{\sqrt 2 r_h}+\frac 12\log 2 -\frac 12$,\footnote{This particular expression of $C_4$ has been chosen for the sake of comparison with literature discussed below.} and $F_{1\mu\nu}=\p_\mu A_{1\nu}-\p_\nu A_{1\mu}$, and similarly for $F_{2\mu\nu}$.

Let us now specialize to the AdS Schwarzschild metric with $d=4$. We find that the coefficients evaluate to
\begin{gather} a_{00}=0,\quad b_{00}=\frac{2\pi} {\beta^2},\quad g_{00}=\frac{2\pi^2}{ \beta^{2}},\quad f_{00}=-\frac{2\pi}\beta,\quad g_{10}=u_{00}=0,\quad v_{00}=-\frac{\pi}{\beta}, \cr
a_{20}=0,\quad a_{02}= 0, 
\quad f_{10}=-\frac{48 G+7\pi^2}{96\pi},\quad b_{20}=\frac\pi4,\quad b_{02}=-\frac\pi8,\quad c_{00}={\pi \ov 8} ,\cr
 g_{20}=0,\quad
 g_{02}=\frac{1-2\log 2}2,\quad h_{00}=\frac{1-\log 2}2-\frac\pi4,\quad v_{10}=\frac{1-\log 2}2,\cr
 u_{10}=\frac{1-\log 2}2+\frac\pi4,\quad  w_{00}=-\frac 12 \ ,
\end{gather}
where $G$ is the Catalan's constant. One can verify that the above satisfies dynamical KMS invariance
\be I_{\text{EFT}}[\tilde B_{1\mu},\tilde B_{2\mu}]=I_{\text{EFT}}[B_{1\mu}, B_{2\mu}]\ ,\ee
where \cite{Glorioso:2018wxw}
\be \tilde B_{1\mu}(x^\mu)=B_{1\mu}(-x^0+i\theta,-x^i),\quad \tilde B_{2\mu}(x^\mu)=B_{2\mu}(-x^0-i(\beta_0-\theta),-x^i)\ .\ee

The expressions of the currents are obtained from $I_{\text{EFT}}$ by varying with respect to $B_{a\mu}$ and $B_{r\mu}$ as in (\ref{Ja}). The conservation equation of $J_a^\mu$ to first order reads, explicitly

\bea\label{eo1}
\p_\mu J^\mu_a={1 \ov Q_0} \p_0 \hat B_{0a} +  K_h \p_0 \p_i \hat B_{ia} = 0 , 
\eea
which, using~\eqref{disp},  can also be written as
\be
\p_0 \rho_c  = 0
\ee
thus confirming the statement that conservation of the horizon charge is equivalent to conservation of the noise current.

Using (4.43) of \cite{CGL} one finds, for the transverse sector,
\bea
G^S_{\al \al} &=&  2 \pi T^2 + {\pi \ov 4} \om^2 - \frac\pi8 q^2 + O(\om^4, q^2, \om^2 q^2) \cr
G^R_{\al \al} &= & i \pi T \om + \frac{1-\log 2}2 \om^2  - \ha q^2  - {i  \pi \beta \ov 16}  \om q^2 + {i \pi \beta \ov 24}  \om^3+
  O(\om^4, q^2 \om^2, q^4)    \ .
\eea
From (4.48) of \cite{CGL} we have, for longitudinal sector
\be \label{pil}
\Pi^L = {- \pi T  + i \frac{1-\log 2}2  \om - \frac{1-2\log 2}{4 \pi T}  q^2 -{\pi \beta \ov 24}  \om^2  +
     \cdots
 \ov i  \om - {q^2 \ov 2 \pi T}   -
 {i \log 2 \ov 4 \pi^2 T^2 } \om q^2  + O(q^4, \om^4, \om^2 q^2) \cdots}
\ee
and
\be\label{gl}
G^L = {2 \pi T^2 - ({\pi \ov 8} + {\log 2 \ov \pi} )  q^2
+ {\pi \ov 4} \om^2
+ O(q^4, \om^4, q^2 \om^2)
 \ov \om^2 +  {q^4 \ov 4 \pi^2 T^2}
 -  {\log 2 \ov 2 \pi^2 T^2} \om^2 q^2  + O(\om^6, \cdots)} \ .
\ee

As a non-trivial check of the construction, one can verify that (\ref{pil}),(\ref{gl}) agree precisely with~\cite{Kovtun:2005ev}, in the low-energy limit $\om, q \ll T$.

\section{Discussion and conclusions} \label{sec:5}

In this paper we proposed a simple prescription for real-time correlation functions defined on a
Schwinger-Keldysh contour for gravity geometries with a dynamical horizon.
We first applied the prescription to a scalar field in an external black brane geometry, showing that it dramatically simplifies the extraction of the long-time and large-distance limit of real-time correlation functions.
We also showed that the same procedure can be straightforwardly generalized to a black hole geometry with a slowly-varying horizon. Finally, we applied the formalism to give a gravity derivation of the non-equilibrium effective field theory of diffusion proposed in~\cite{CGL} to quadratic order in deviations from equilibrium and second order in derivatives.
The derivation contains a few new elements compared with calculation of real-time correlation functions, as one needs to keep
the hydrodynamic fields for diffusion off-shell. We showed that this can be achieved by not imposing the Gauss Law and
imposing a horizon boundary condition.

There are a number of immediate future directions. Firstly, one should study how the prescription works in the
calculation of higher-point real-time correlation functions. We saw that the bulk fields exhibit logarithmic behavior near the horizon. It is important to understand whether such behavior leads to complications in the presence of nonlinearity.
Secondly, it would be interesting to perform a more systematic study of correlation functions for non-equilibrium geometries with a dynamical horizon.\footnote{The analytic procedure requires the metric components to be analytic functions of the radial coordinate.} This includes generalizing the discussion of Sec.~\ref{sec:3b} to higher derivative orders and studying more general time-dependent geometries. Such investigations may lead to a more geometric way to identify local temperature of a slowly-varying black hole horizon which we initiated in Sec.~\ref{sec:3b}.
 Finally,  it should be possible to generalize the derivation of the effective action for diffusion to the
effective action for the full dissipative hydrodynamics (see~\cite{Grozdanov:2013dba,CGL,Haehl:2015uoc,GaoL,CGL1,yarom,Glorioso:2018wxw,Haehl:2018lcu,Jensen:2018hse} for discussions of dissipative hydrodynamic actions). This requires dealing with full nonlinear bulk gravity.

\vspace{0.2in}   \centerline{\bf{Acknowledgements}} \vspace{0.2in}
We thank Yifan Wang for collaboration at the initial stage of the work and Julian Sonner for discussions and in particular questions which motivated the discussion of Sec.~\ref{sec:3b}, and for comments on the draft.
This work is supported by the Office of High Energy Physics of U.S. Department of Energy under grant Contract Number  DE-SC0012567.  P. G. was supported by a Leo Kadanoff Fellowship.

\end{document}